\documentclass{jetpl}
\twocolumn
\usepackage{graphicx}
\usepackage{bm}
\lat

\title{Phonon-particle coupling effects in odd-even double  mass differences of magic nuclei.}%

\rtitle{Phonon-particle coupling effects in odd-even double  mass differences of magic nuclei.}%

\sodtitle{Phonon-particle coupling effects in odd-even double  mass differences of magic nuclei.}%

\author{
 E.\,E.\,Saperstein$^{*,**}$\/\thanks{e-mail: saper@mbslab.kiae.ru},
 M.\,Baldo$^{\dagger}$,
 N.\,V.\,Gnezdilov$^{+}$,
 S.\,V.\,Tolokonnikov$^{*,***}$}

\rauthor{E.\,E.\,Saperstein, M.\,Baldo, N.\,V.\,Gnezdilov, S.\,V.\,Tolokonnikov}

\sodauthor{E.\,E.\,Saperstein, M.\,Baldo, N.\,V.\,Gnezdilov, S.\,V.\,Tolokonnikov}

\address{$^{*}$National Research Centre Kurchatov Institute, pl.
Akademika Kurchatova 1, Moscow, 123182 Russia\\
$^{**}$National Research Nuclear University MEPhI, 115409 Moscow, Russia\\
$^{\dagger}$INFN, Sezione di Catania, 64 Via S.-Sofia, I-95125 Catania, Italy\\
$^{+}$Instituut-Lorentz, Universiteit Leiden, P.O. Box 9506, 2300 RA Leiden, The Netherlands\\
$^{***}$Moscow Institute of Physics and Technology, 141700 Dolgoprudny, Russia}

\dates{\today}{*}

\abstract{ A method is developed to consider the particle-phonon coupling (PC) effects in the problem
of finding odd-even double mass differences (DMD) of magic nuclei within the approach starting from
the free $NN$-potential.  Three PC effects are taken into account, the phonon induced interaction, the
renormalization of the ``ends'' due to the $Z$-factors and the change of the single-particle energies.
We use the perturbation theory in $g^2_L$, where $g_L$ is the vertex of the $L$-phonon creation. PC
corrections to single-particle energies are found self-consistently with an approximate account for
the so-called tadpole diagram.   Results for double-magic $^{132}$Sn and  $^{208}$Pb nuclei show that
the PC corrections make agreement with the experimental data better.}


\begin{document}

\newcommand{\beq}{\begin{equation}}
\newcommand{\eeq}{\end{equation}}
\newcommand{\bea}{\begin{eqnarray}}
\newcommand{\eea}{\end{eqnarray}}
\newcommand{\eps}{\varepsilon}
\newcommand{\bfg}{\boldsymbol}

\maketitle

Recently, the semi-microscopic model developed first for the pairing problem \cite{Pankr1,Pankr2,Sap1}
was successfully applied to the problem of finding the odd-even double mass differences (DMD) of magic
nuclei \cite{Gnezd1,Gnezd2,Gnezd3}. In the pairing problem, this model starts from the Brueckner
theory which results in the BCS gap equation \beq  \Delta = {\cal V} G G^s, \label{BCS}\eeq where
${\cal V}$ is a ``realistic'' $NN$-potential (the Argonne ${\rm v}_{18}$  in our case), and $G$
($G^s$) is the one-particle Green function without (with) pairing. In the case of direct solving this
equation in a single-particle basis \cite{milan1,milan2,milan3} for the $^{120}$Sn nucleus, a serious
problem of slow convergency exists. To overcome this problem, a two-step renormalization method of
solving the gap equation was used in Refs. \cite{Pankr1,Pankr2,Sap1}.   The complete Hilbert space of
the pairing problem $S$ is split in the model subspace $S_0$, including the single-particle states
with energies less than a separation energy $E_0$, and the complementary one, $S'$. The gap equation
is solved in the model space with the effective pairing interaction (EPI) obeying the Bethe--Goldstone
type equation in the subsidiary space:\beq {\cal V}_{\rm eff} = {\cal V}  + {\cal V}  G  G {\cal
V}_{\rm eff}|_{S'}. \label{Vef} \eeq In these calculations, the energy density functional (EDF) by
Fayans et al. \cite{Fay1,Fay4,Fay5,Fay} was used, which is characterized by the bare mass, $m^*{=}m$.
The set DF3 \cite{Fay4,Fay} of the EDF parameters and its modifica- \vskip 0.5cm \noindent tion DF3-a
\cite{Tol-Sap} were employed.

In contrast, in Refs. \cite{milan2,milan3} an essentially non-bare effective mass  of the
Skyrme--Hartree--Fock method  (the SLy4 EDF \cite{SLy4}) was used with a dramatic suppression of the
gap $\Delta$ values. To obtain a result close to the experimental value $\Delta_{\rm exp}\simeq
1.3\;$MeV, the particle-phonon coupling (PC)  corrections to the BCS approximation were introduced. In
addition, the contribution of the induced interaction due to exchange of high-lying collective
excitations was included in \cite{milan3}. High uncertainties in a direct finding of all these
corrections to the simplest BCS scheme with bare nucleon mass were discussed in detail in
\cite{Bald1,BCS50}.

The scale of these uncertainties grow with appearance of the results obtained by Duguet et al.
\cite{Dug1,Dug2} for a number of nuclei with the use of the ``low-k'' force ${\cal V}_{\rm low-k}$
\cite{Kuo,Kuo-Br} which is rather soft. The quasi-potential ${\cal V}_{\rm low-k}$ is defined in such
a way that it describes the $NN$-scattering phase shifts at momenta $k{<}\Lambda$, where $\Lambda$  is
a parameter corresponding to the limiting energy  $\simeq 300\;$MeV, which is much less than the value
of $E_{\max}{=} 800\;$MeV in \cite{milan2,milan3} and helps to carry out systematic calculations.  The
force ${\cal V}_{\rm low-k}$ vanishes  for $k{>}\Lambda$, so that in the gap equation one can restrict
the energy range
 to $E_{\max} {\simeq} 300\;$MeV. Usually the low-k force is
found starting from some realistic $NN$-potential ${\cal V}$ with
the help of the Renormalization Group method, and the result does
not practically depend on the particular choice  of ${\cal V}$
\cite{Kuo}. In addition, in Ref. \cite{Dug1} ${\cal V}_{\rm low-k}$
was found starting from the Argonne potential v$_{18}$, that is
different only a little from Argonne v$_{14}$, used in Ref.
\cite{milan3}. Finally, in Ref. \cite{Dug1} the same SLy4
self-consistent basis was used as in Ref. \cite{milan3}. Thus, the
inputs of the two calculations look very similar, but the results
turned out to be strongly different. In fact, in Ref. \cite{Dug1}
the value $\Delta_{\rm BCS}\simeq 1.6\;$MeV was obtained for the
same nucleus $^{120}$Sn which is already bigger than the
experimental one by $\simeq 0.3\;$MeV. In Refs. \cite{Pankr1,Bald1,BCS50}
the reasons of these contradictions were analyzed.  It turned out
that these two calculations differ in the way they take
into account the effective mass. It implies that the gap $\Delta$
depends not only on the value of the effective mass at the Fermi
surface, as it follows from the well-known BCS exponential formula for the
gap, but also on the behavior of the function $m^*(k)$ in a wide
momentum range. However, this quantity is not known sufficiently  well.
An additional problem was specified in Ref. \cite{Dug3} where it was
found that the inclusion of the 3-body force following
from the chiral theory \cite{Epel}  suppresses the gap values much lower than the
experimental ones.

To avoid uncertainties under  discussion, the semi-microscopic model was suggested
\cite{Pankr1,Pankr2,Sap1} in which the EPI (\ref{Vef}) is supplemented with a phenomenological
$\delta$-function addendum: \bea {\cal V}_{\rm eff}({\bf r}_1,{\bf r}_2,{\bf r}_3,{\bf r}_4) = V^{\rm
BCS}_{\rm eff}({\bf r}_1,{\bf r}_2,{\bf r}_3,{\bf r}_4) + \nonumber \\
\gamma C_0 \frac {\rho(r_1)}{\bar{\rho}(0)}
\prod_{i=2}^4\delta ({\bf r}_1 - {\bf r}_i). \label{Vef1} \eea Here
$\rho(r)$ is the density of nucleons of the kind under
consideration, and $\gamma$ are dimensionless
phenomenological parameters. The average central
density ${\bar{\rho}(0)}$ in the denominator of the
additional term is obtained with averaging the density $\rho(r)$   over the interval of $r{<}2\;$fm.

The odd-even DMD we deal are defined in terms of nuclear masses $M(N,Z)$ as follows: \beq
D_{2n}^+(N,Z) {=}  M(N{+}2,Z){+} M(N,Z){-}2M(N{+}1,Z),\label{d2npl} \eeq \beq D_{2n}^-(N,Z) {=}
{-}M(N-2,Z){-} M(N,Z){+}2M(N{-}1,Z),\label{d2nmi}\eeq \beq D_{2p}^+(N,Z) {=}  M(N,Z{+}2){+}
M(N,Z){-}2M(N,Z{+}1),\label{d2ppl} \eeq \beq D_{2p}^-(N,Z) {=} {-}M(N,Z-2){-}M(N,Z){+}
2M(N,Z{-}1).\label{d2pmi}\eeq The ``experimental'' gap values $\Delta_{\rm exp}$ we mentioned above
are usually identified with a half of their value.

In magic nuclei which are non-superfluid, these odd-even mass DMD (\ref{d2npl})--(\ref{d2pmi}) can be
expressed in terms of the same EPI (\ref{Vef}) as the pairing gap \cite{Gnezd1,Gnezd2,Gnezd3}. It can
be easily proved starting from the Lehmann expansion  for the two-particle Green function $K$ in a
non-superfluid system. In the single-particle wave functions $|1\rangle{=}|n_1,l_1,j_1,m_1\rangle$
representation, it  reads \cite{AB}: \beq K_{12}^{34}(E)=\sum_s \frac {\chi^s_{12}\chi^{s+}_{34}}
{E-E_s^{+,-} \pm i\gamma}, \label{Lem}\eeq where $E$ is the total energy in the two-particle channel
and $E_s^{+,-}$ denote the eigen-energies of  nuclei with
 two particles and two holes, respectively, added to the original nucleus.  Instead of the Green function $K$, it is convenient
to use the two-particle interaction amplitude
$\Gamma$: \beq K = K_0 + K_0 \Gamma K_0, \label{gam}\eeq where
$K_0=GG$. Within the Brueckner theory, the amplitude
$\Gamma$ obeys the following equation \cite{AB}: \beq \Gamma = {\cal V}+{\cal V} GG
\Gamma, \label{eqgam}\eeq where ${\cal V}$ is the same $NN$-potential as in Eq. (\ref{BCS}),
 which does not depend on the energy. Then the integration over the relative energy can be readily carried
out in Eq. (\ref{eqgam}): \beq
 A_{12} {=}  \int \frac {d\eps}{2\pi
i}G_1\left(\frac E 2 {+}\eps \right) G_2\left(\frac E 2 {-}\eps
\right)
 {=}\frac {1{-}n_1{-}n_2}
{E{-}\eps_1{-}\eps_2}, \label{Alam} \eeq where $\eps_{1,2}$ are the single-particle energies and
$n_{1,2}{=}(0;1)$, the corresponding occupation numbers. As a consequence, Eq. (\ref{eqgam}) reduces
to the following form: \beq \Gamma = {\cal V}+{\cal V} A \Gamma. \label{eqgam1}\eeq

The two-particle amplitude $\Gamma(E)$ possesses the same poles $E_s^{+,-}$ as the Green function $K$.
After simple manipulations \cite{Gnezd1}, one can obtain the equation for the eigenfunctions $\chi^s$:
\beq (E_s-\eps_1-\eps_2) \chi^s_{12}=(1-n_1-n_2) \sum_{34}{\cal V}_{12}^{34}
\chi^s_{34}\label{eqchi}.\eeq It is different from the Shr\"{o}dinger equation for two interacting
particles in an external field only for the factor $(1-n_1-n_2)$ which reflects the many-body
character of the problem, in particular, the Pauli principle. As in the pairing problem, the angular
momenta of two-particle states  $|12\rangle$, $|34\rangle$ are coupled to the total angular momentum
$I{=}0$ ($S{=}0, L{=}0$).

The relevance of the same interaction ${\cal V}_{\rm eff}$  for these two different problems agrees with the well-known theorem by Thouless  \cite{Thouless} stating that the gap equation reduces to the in-medium Bethe-Salpeter equation provided the gap $\Delta$ vanishes. In our case, the homogeneous counterpart of Eq. (\ref{eqgam1}) is the Bethe-Salpeter equation under discussion, and the Shr\"{o}dinger-like  equation (\ref{eqchi}) can be obtained from it with the usual procedure. In nuclear physics, this point was evidently first discussed in \cite{Sap-Tr}, where the DMD values for double-magic nuclei were analyzed
within the theory of finite Fermi systems \cite{AB}. In this article,
the density dependent EPI was introduced and arguments were found in favor
of the surface dominance in this interaction.

The direct solution of this equation is  complicated by the same reasons as for the BCS gap equation
described above. The same two-step method is used in combination with LPA to overcome this difficulty.
As a consequence,  Eq. (\ref{eqchi}) is transformed into the analogous equation in the model
space:\beq (E_s{-}\eps_1{-}\eps_2) \chi^s_{12}{=}(1{-}n_1{-}n_2) {\sum_{34}}^0 \left({\cal V}_{\rm
eff}\right)_{12}^{34}\; \chi^s_{34},\label{eqchi0}\eeq where the effective interaction ${\cal V}_{\rm
eff}$ coincides with that of the pairing problem, Eq. (\ref{Vef}), provided the same value of the
separation energy $E_0$ is used. The next step consists in the use of the ansatz (\ref{Vef1}) to take
into account corrections to the Brueckner theory with a phenomenological addendum ($ \sim \gamma$).
These corrections are obviously the same as discussed above for the BCS theory.  In Refs.
\cite{Gnezd1,Gnezd2,Gnezd3}, the semi-microscopic model was successfully applied to non-superfluid
components of semi-magic nuclei with the same value of $\gamma{=}0.06$ as for the pairing gap.

\begin{figure}
\centerline {\includegraphics [width=60mm]{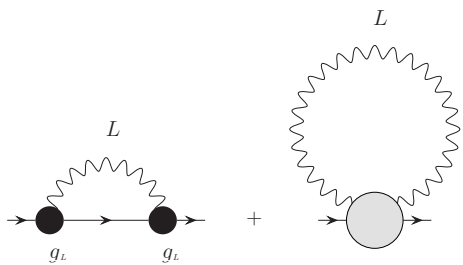}} \caption{Fig. 1. PC corrections
to the mass operator. The gray blob denotes the ``tadpole'' term.} \label{fig:PC3}
\end{figure}

\begin{figure}
\centerline {\includegraphics [width=80mm]{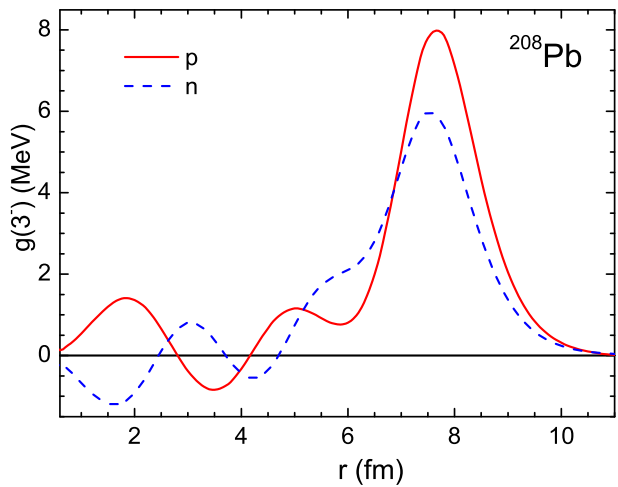}} \caption{Fig. 2. The vertex $g_L$ for the
$3^-_1$ state in $^{208}$Pb.} \label{fig:Pb3-}
\end{figure}

 \begin{figure}
\centerline {\includegraphics [width=12mm]{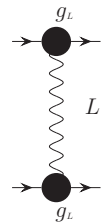}} \caption{Fig. 3. The
phonon induced interaction.} \label{fig:PC1}
\end{figure}

\begin{figure}
\centerline {\includegraphics [width=50mm]{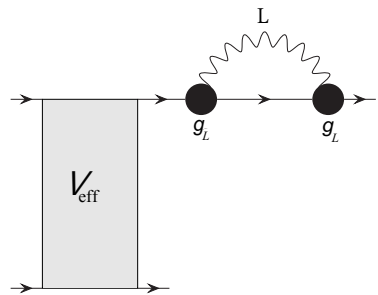}} \caption{Fig. 4. An example of the PC ``end''
correction.} \label{fig:PC2}
\end{figure}

In this work, we develop a method of direct account for the PC corrections to the DMD values, together
with possible change of the optimal value of $\gamma$. The introduction of the PC corrections to Eq.
(\ref{eqchi0}) consists, first, of the change of $\eps_{\lambda}$ on the l.h.s. to
$\widetilde{\eps}_{\lambda}{=}\eps_{\lambda}{+}\delta \eps_{\lambda}^{\rm PC}$ and, second, a similar
change of the ${\cal V}_{\rm eff}$ quantity on the r.h.s., to $\widetilde{\cal V}_{\rm eff}$, with the
same meaning of the ``tilde'' symbol. The explicit form of this PC corrected equation reads: \beq
(E_s{-}\widetilde{\eps}_1{-}\widetilde{\eps}_2) \chi^s_{12}{=}(1{-}n_1{-}n_2) {\sum_{34}}^0
\left(\widetilde{\cal V}_{\rm eff}\right)_{12}^{34}\; \chi^s_{34},\label{eqPCchi0}\eeq

Let us begin with single-particle energies. We follow here the method developed in \cite{levels}. Note
also that recently PC corrections to the single-particle energies within different self-consistent
approaches were studied in Refs. \cite{Litv-Ring,Bort,Dobaczewski,Baldo-PC}. To find the
single-particle energies with account for the PC effects, we solve the following equation: \beq
\left(\eps-H_0 -\delta \Sigma^{\rm PC}(\eps) \right) \phi =0, \label{sp-eq}\eeq where $H_0$ is the
quasiparticle Hamiltonian with the spectrum $\eps_{\lambda}^{(0)}$ and $\delta \Sigma^{\rm PC}$ is the
PC correction to the quasiparticle mass operator. After expanding this term in the vicinity of
$\eps=\eps_{\lambda}^{(0)}$ one finds \beq \eps_{\lambda}=\eps_{\lambda}^{(0)} + Z_{\lambda}^{\rm PC}
\delta \Sigma^{\rm PC}_{\lambda\lambda}(\eps_{\lambda}^{(0)}) ,\label{eps-PC}\eeq with obvious
notation. Here $Z^{\rm PC}$ denotes the $Z$-factor due to the PC effects,


 \beq Z_{\lambda}^{\rm PC} =\left({1- \left(\frac {\partial} {\partial \eps}
 \delta \Sigma^{\rm PC}(\eps) \right)_{\eps=\eps_{\lambda}^{(0)}}}\right)^{-1}. \label{Z-fac}\eeq

Expression (\ref{eps-PC}) corresponds to the perturbation
theory in the $\delta \Sigma$ operator with respect to $H_0$. In this
article, we limit ourselves to magic nuclei where the so-called
$g_L^2$-approximation, $g_L$ being the $L$-phonon creation
amplitude, is, as a rule, valid. It is worth mentioning that Eq.
(\ref{eps-PC}) is more general, including, e.g., $g_L^4$ terms. In the case when several $L$-phonons are taken into account,
the total PC variation of the mass operator in Eqs. (\ref{sp-eq})--(\ref{Z-fac}) is just the sum:
\beq \delta \Sigma^{\rm PC} = \sum_L \Sigma^{\rm PC}_L . \label{sum-L}\eeq

The diagrams for the $\delta \Sigma^{\rm PC}_L$ operator within the $g_L^2$-approximation are displayed in Fig. 1.
The first one is the usual pole diagram, with obvious notation, whereas the second, ``tadpole'' diagram represents the sum
of all non-pole diagrams of the $g_L^2$ order.

In the obvious symbolic notation, the pole diagram corresponds to
$\delta\Sigma^{\rm pole}=(g_L,D_LGg_L)$  where $D_L(\omega)$ is the
phonon $D$-function.  Explicit expression for the pole term is as follows:
\bea \delta\Sigma^{\rm
pole}_{\lambda\lambda}(\epsilon)&=&\sum_{\lambda_1\,M}
|\langle\lambda_1|g_{LM}|\lambda\rangle|^2 \nonumber\\
&\times&\left(\frac{n_{\lambda_1}}{\eps+\omega_L-
\eps_{\lambda_1}}+\frac{1-n_{\lambda_1}}{\eps-\omega_L -\eps_{\lambda_1}}\right), \label{dSig2} \eea
where $\omega_L$ is the excitation energy of the $L$-phonon.
In the coordinate form of
their creation amplitudes $g_L({\bf r})$ the surface peak dominates,
\beq g_L(r)=\alpha_L \frac {dU} {dr} +\chi_L(r), \label{gLon}\eeq
where $U(r)$ is the nuclear mean-field potential, and the in-volume correction $\chi_L(r)$ being rather small. In Fig. 2, it is illustrated for the $3^-_1$-state in $^{208}$Pb.
If one neglects in-volume contributions, the tadpole PC term is reduced \cite{levels}
to a simple form:
\beq
\delta\Sigma^{\rm tad}_L = \frac {\alpha_L ^2} 2 \frac {2L+1} 3
\triangle U(r). \label{tad-L}\eeq  In this work, following to \cite{levels}, we use this approximation.

\begin{table*} \caption{Table 1. Different PC corrections to odd-even double mass  differences of magic nuclei.\label{tab:delPCD2pm}}
\begin{center}
\begin{tabular}{|c|c|c|c|c| c|c|c|l|}
\hline\noalign{\smallskip}
&   & $D_2^{(0)}$  & $\delta D_2 (Z^{\rm PC})$
  & $\delta D_2 ({\cal V}_{\rm ind}^{\rm PC})$    &  $\delta D_2(\delta \eps^{\rm PC})$ & $\delta D_2^{\rm PC}$ & $D_2^{\rm PC}$ &   $D_2^{\rm exp}$   \\
\hline

$^{132}$Sn-pp  & $D_2^-$&     3.184&  -1.506  & -0.015   & -0.982  & -1.198      &1.986    &    2.027(160)\\
               & $D_2^+$&    -2.763&   1.319  & -0.250   & 1.710   & 1.494       &   -1.269&   -1.234(6) \\

 $^{132}$Sn-nn  & $D_2^-$&    2.301&  -0.396   &  0.369   &  -0.009   & -0.161   & 2.140&       2.132(9)\\
                & $D_2^+$&   -1.165&  0.217    &  -0.102  &  -0.045   &  0.094   & -1.071&     -1.227(6)    \\

$^{208}$Pb-pp  & $D_2^-$&   1.680&   -0.824   &   -0.083  &   0.569  & -0.745    &         0.935&    0.627(22)\\
               & $D_2^-$&  -2.286&  1.049     &  -0.167   &  -0.329  &  0.830   &        -1.456&     -1.1845(11)  \\

$^{208}$Pb-nn  & $D_2^-$&   0.778&  -0.275   &   0.174  &   0.205   & -0.113   &    0.665&       0.63009(11)\\
               & $D_2^-$&  -1.156&   0.443   &  -0.691  &  -0.021   & 0.165    &   -0.991&      -1.2478(17)\\
\hline

\end{tabular}
\end{center}
\end{table*}

\begin{table*} \caption{Table 2. Difference $\delta D_2$ between theoretical and experimental values of DMD for different versions of the theory.\label{tab:delTHEORY}}
\begin{center}
\begin{tabular}{|c|c|c|c|c|c|l|}
\hline\noalign{\smallskip}
& $\gamma{=}0$   & $\gamma{=}0.06$ & $(\gamma{=}0.06)^{\rm PC}$ & $(\gamma{=}0.03)^{\rm PC}$  &  $(\gamma{=}0)^{\rm PC}$  & $D_2^{\rm exp}$ \\
\hline\noalign{\smallskip}

$^{132}$Sn-pp  & -1.529&    -0.641&      -0.210&    -0.117&   -0.035&     -1.234(6)\\

$^{132}$Sn-nn  &  0.169&    -0.390&      -0.440&    -0.231&    0.008&     2.132(9) \\
               &  0.062&     0.327&       0.348&     0.260&    0.156&   -1.227(6)\\

$^{208}$Pb-pp  &  1.053&     0.373&       0.091&     0.188&    0.308&    0.627(22)\\
               & -1.101&    -0.282&       0.065&    -0.091&   -0.271&   -1.1845(11) \\

$^{208}$Pb-nn  &  0.148&    -0.100&      -0.136&    -0.060&    0.035&    0.63009(11) \\
               &  0.092&     0.427&       0.428&     0.349&    0.257&   -1.2478(17)\\
\hline
$\langle\delta D_2\rangle_{\rm rms}$ & 0.82138  &   0.39298  &  0.28605  &   0.20827 &   0.19323 \\
\noalign{\smallskip}\hline

\end{tabular}
\end{center}
\end{table*}

The tadpole term does not depend on the energy, therefore the $Z^{\rm PC}$-factor (\ref{Z-fac}) is
determined with the pole term only and can be found directly in terms of the energy derivative of Eq.
(\ref{dSig2}).

Let us go to PC corrections to the r.h.s. of Eq. (\ref{eqchi0}). They include the phonon induced
interaction, Fig. \ref{fig:PC1}, and the ``end corrections''. An example of them is given in Fig.
\ref{fig:PC2}. Partial summation of such diagrams results in the ``renormalization'' of ends: \beq
|\lambda\rangle \to |\widetilde{\lambda}\rangle =\sqrt{Z_{\lambda}^{\rm PC}} |\lambda\rangle.
\label{ren_end}  \eeq In the result, we get \bea \langle 1 1'|\widetilde{\cal V}_{\rm eff}|22'\rangle
&=& \sqrt{Z_1^{\rm PC}Z_{1'}^{\rm PC}Z_2^{\rm PC}Z_{2'}^{\rm PC}}
\nonumber \\
&\times& \langle 1 1'|{\cal V}_{\rm eff} +  {\cal V}_{\rm ind}|22'\rangle. \label{Vtild}\eea

Remind that we deal with the channel with $I{=0},S{=0},L{=0}$. Hence, the states $i,i'$ in
(\ref{Vtild}) possess the same single-particle angular momenta,
$j_1{=}j_{1'},l_1{=}l_{1'};j_2{=}j_{2'},l_2{=}l_{2'}$. In this case, the explicit expression of the
matrix element of ${\cal V}_{\rm ind}$ is as follows:
\bea \langle 1 1'|{\cal V}_{\rm ind}|22'\rangle = - \frac {2 \omega_L} {\sqrt{(2j_1+1)(2j_2+1)}}  \nonumber \\
\times \frac {\bigl(\langle j_1 l_1|| Y_L || j_1 l_1\rangle (g_L)_{11'}\bigr) \bigl(\langle j_2 l_2||
Y_L || j_2 l_2\rangle (g_L)_{22'}\bigr)^* }{\omega_L^2-(\eps_2-\eps_1)^2 }, \label{Vind} \eea where
$\langle \;|| Y_L || \;\rangle$ stands for the reduced matrix element \cite{BM1}, and $(g_L)_{ii'}$
are the radial matrix elements of the vertex $g_L(r)$.

The above formulas (\ref{sp-eq})--(\ref{Vind}) were used to find from Eq. (\ref{eqPCchi0}) the PC
corrections to the odd-even DMD values for double-magic nuclei $^{132}$Sn and $^{208}$Pb. The Fayans
EDF DF3-a \cite{Tol-Sap} was used which reproduces characteristics of the $L$-phonons in these nuclei
sufficiently well \cite{levels}. As it is well known,  PC corrections are important mainly for
single-particle states  close to the Fermi surface. In practice, we solve the PC corrected equation
(\ref{eqPCchi0}) limiting ourselves with two shells nearby the Fermi level. In Table 1, the effect of
each PC correction to a DMD value is given separately. In this set of calculations we put $\gamma{=}0$
in Eq. (\ref{Vef1}) which determines the EPI of the semi-microscopic model, hence $D_2^{(0)}$ means
the direct prediction for the DMD of the Brueckner theory. The next columns present separate PC
corrections to this quantity. So, the 2-nd column shows the result of application of Eq. (\ref{Vtild})
with ${\cal V}_{\rm ind}{=}0$, whereas the 3-rd one presents the effect of ${\cal V}_{\rm ind}$ itself
with $Z_1^{\rm PC}{=}...{=}Z_{2'}^{\rm PC}{=}1$. The   column 4 shows the effect of PC corrections to
the single-particle energies in Eq. (\ref{eqPCchi0}) only. At last, column 5 presents the total PC
effect $\delta D_2^{\rm PC}{=}D_2^{\rm PC}-D_2^{(0)}$, where $D_2^{\rm PC}$ (column 6) is the solution
of Eq. (\ref{eqPCchi0}) with all PC corrections included. As it  should be, the value of $\delta
D_2^{\rm PC}$ does not equal the sum of the values in previous three columns because of an
interference between different PC effects. Experimental DMD values are found from the mass table
\cite{mass}.

The $Z$-factor effect (column 2)  always has the sign opposite to that of $D_2^{(0)}$ value thus
suppressing the absolute value of $D_2^{(0)}$. This is a trivial consequence of the $Z^{\rm PC}<1$
condition.  The scale of the suppression varies from $\simeq 20$\% (the neutron $D_2^+$ mode in
$^{132}$Sn) to $\simeq 50$\% (both proton modes in $^{132}$Sn and $^{208}$Pb). It agrees with average
values of the $Z^{\rm PC}$-factors,  $Z^{\rm PC}_{\lambda}\simeq (0.7 \div 0.9)$, of these nuclei
found in \cite{levels} or  \cite{spectr1,spectr2}. In all cases where the PC effect due to the induced
interaction (column 3) is big, its sign coincides with that of $D_2^{(0)}$, i.e. it corresponds to an
additional attraction. Two exceptions, the proton $D_2^-$ mode in both nuclei, occur in the cases of
very small value of this effect, much less than that due to the $Z$-factor. At last, go to the
single-particle energy effect (column 4). Here there are five cases where this effect is rather big
and three, where it is negligible. In all the cases of the first part, this effect helps to make
agreement with the data better. The total PC correction (column 5) has always the correct sign with
one exception, the neutron $D_2^-$ mode in $^{208}$Pb. Fortunately, in this ``bad'' case the PC
correction is not big and spoils agreement not much. On the contrary, in many ``good'' cases this
correction is large and helps to improve the initial $D_2^{(0)}$ value significantly. In all the
cases, the PC effect results in a suppression of the initial DMD value, i.e. it acts qualitatively as
the phenomenological term in Eq. (\ref{Vef1}) for the EPI of the semi-phenomenological model. This
makes it reasonable to try to search a new optimal value of the parameter $\gamma$ with account for
the PC effects.

Results of such attempt are given in table 2. To make the comparison with experiment more transparent,
we present  differences between each theoretical prediction and the corresponding experimental value.
We exclude from the analysis one case, the proton $D_2^-$ mode in $^{132}$Sn, where the experimental
datum does not possess  sufficiently high accuracy. In the last line, we put the rms deviation of each
version of the theory from the data. Of course such average is not so much indicative for so small
number of averaged quantities, nevertheless it helps to feel a tendency.   The column  2 corresponding
to $\gamma{=}0.06$ without PC corrections has, of course, better accuracy than the column 1
corresponding to the pure Brueckner theory. However, it gives way to all three next columns
corresponding different values of $\gamma$ with PC corrections. It is difficult to choose between two
columns, 4 and 5 with $\gamma{=}0.03$ and $\gamma{=}0$ correspondingly, but it looks highly believable
that the initial value $\gamma{=}0.06$ of the semi-microscopic model should be taken smaller after
explicit inclusion of the  PC corrections.

A wider amount of nuclei should be analyzed with PC corrections included for more definite conclusions
on the optimal value of the phenomenological parameter $\gamma$ of the semi-microscopic model. It can
include other magic nuclei and non-superfluid subsystems of semi-magic nuclei as well. However, a
careful choice should be made of nuclei where the perturbation theory in the PC coupling vertex is
valid.

\vskip 1cm The work was partly supported  by the Grant NSh-932.2014.2 of the Russian Ministry for
Science and Education, and by the RFBR Grants 13-02-00085-a, 13-02-12106-ofi\_m, 14-02-00107-a,
14-22-03040-ofi\_m. Calculations were partially carried out on the Computer Center of Kurchatov
Institute. E. S. thanks the INFN, Seczione di Catania, for hospitality.

{}

\end{document}